# Robust spin-orbit coupling in semi-metallic SrIrO$_3$ under hydrostatic pressure

D. Fuchs,[1,*] A. K. Jaiswal,[1] F. Wilhelm,[2] D. Wang,[3] A. Rogalev,[2] and M. Le Tacon[1]

[1]Karlsruhe Institute of Technology, Institute for Quantum Materials and Technologies, Kaiserstraße 12, 76131 Karlsruhe, Germany;
[2]ESRF-European Synchrotron Radiation Facility, 38043 Grenoble, France
[3]Karlsruhe Institute of Technology, Institute of Nanotechnology and Karlsruhe Nano Micro Facility (KNMFi), Kaiserstraße 12, 76131 Karlsruhe, Germany;

*corresponding author: dirk.fuchs@kit.edu

The semi-metallic behavior of the perovskite iridate SrIrO$_3$ shifts the end-member of the strongly spin-orbit (SO) coupled Ruddlesden-Popper series Sr$_{n+1}$Ir$_n$O$_{3n+1}$ away from the Mott insulating regime and the half-filled pseudospin $J_{eff}$=1/2 ground state well-established in the layered iridates ($n$ = 1 and 2). To investigate the robustness of the SO coupled ground state of SrIrO$_3$, X-ray absorption spectroscopy was carried out at the Ir $L_{2,3}$ edges under hydrostatic pressure up to 50 GPa at room temperature. The effective SO coupling was deduced from the branching ratio of the Ir $L_2$ and $L_3$ white lines. With increasing pressure, the branching ratio decreases, and the Ir $L_{2,3}$ peak positions shift to higher energies. The number of 5$d$ holes remains constant indicating that the spectral weight redistribution and peak shifts arise from orbital mixing between $t_{2g}$ and $e_g$ related states. The expectation value of the angular part of the SO operator <LS> decreases by about 15% at 50 GPa. This reduction, which is very similar to that observed in the layered iridates, is well explained by an increase of the octahedral crystal field due to the shortening of the Ir-O bond-length under compression. Consistent with theoretical predictions, the orbital mixing and <LS> decrease as the crystal field increases. However, the effective SO coupling remains robust against pressure and does not indicate a covalency-driven breakdown within the investigated pressure range.

I. Introduction

When the spin and angular momentum of electrons are strongly coupled due to the relativistic spin-orbit (SO) interaction, the total angular momentum $J$ becomes the relevant quantum number for describing the system. As the strength of spin-orbit coupling (SOC) increases with the atomic number Z, it emerges as a significant energy scale in the low energy physics of 4$d$ and 5$d$ transition metals. There, SOC can give rise to exotic electronic and magnetic quantum states with nontrivial band topologies, including spin liquids [1,2], topological Mott insulators [3], Weyl semi-metals, and axion insulators [4].

In the 5$d$ transition metal oxides Sr$_{n+1}$Ir$_n$O$_{3n+1}$, the SOC strength becomes comparable to the Coulomb interaction and the crystal electric field (CF) splitting. Consequently, the ground state of the Ir$^{4+}$ ion in the cubic CF (10$Dq$) of the IrO$_6$ octahedra is often well described by a pseudospin $J_{eff}$ = ½ picture [5–8]. For $n$ = 1 or 2 layered members of this series, electron correlation and magnetic exchange interactions result in an antiferromagnetic (AFM) ground state. This leads to a splitting of the $J_{eff}$ = ½ state into lower and upper Hubbard bands, producing a "spin-orbital Mott insulator" [7,8]. In contrast, the end member of the Ruddlesden Popper series SrIrO$_3$ (SIO) ($n$ = ∞) exhibits metallic behavior [9,10]. In the conventional framework, the increased coordination and dimensionality of the IrO$_2$ layers are expected to broaden the electronic bands, leading to greater band hybridization compared to the layered compounds. However, photoemission spectroscopy reveals a surprisingly narrow bandwidth in SIO (≈ 0.3 eV), comparable to that of $Sr_2IrO_4$ (0.3-0.8 eV) [11]. This has been attributed to the pronounced orthorhombic distortion and rotation of the $IrO_6$ octahedra in SIO. Alternatively, it has been proposed that the metallic behavior of SIO arises from the absence of AFM order and a weaker electron correlation (U), which, in the layered iridates, is sufficient to open a Mott-Hubbard gap even for small U [12]. In contrast to the 'Mott scenario', where SIO is described by a single half-filled band, this perspective suggests the presence of multiple partially occupied bands with mixed orbital character.

A highly effective technique to probe the SO coupled ground state of iridates is X-ray absorption near-edge spectroscopy (XANES) at Ir $L_{2,3}$ edge. This approach, initially proposed by van der Laan and Thole, relates the intensity ratio $I_{L_3}/I_{L_2}$ of the $L_3$ and the $L_2$ white lines (WL) which are split by SOC to the expectation value of the angular part of the SO operator $<LS>$ [13–15]. According to the SO sum rules, the branching ratio (BR*) for the $p \rightarrow d$ transitions is given by: $BR^* = I_{L_3}/I_{L_2} = (2+r)/(1-r)$, where $r = \langle LS \rangle / \langle n_h \rangle$ and $\langle n_h \rangle$ is the average number of the $5d$ holes [13].

XANES studies on various Ir-based $5d$ compounds, including $Sr_2IrO_4$, $IrO_2$, or $IrCl_3$, reveal strong SOC that are largely independent of chemical composition, crystal structure, or electronic state [16,17]. These materials exhibit a branching ration $BR^*$ of about 7, significantly higher than the statistical value of 2, which corresponds to system with $<LS>$ =0 and is solely determined by the $p_{1/2}$ and $p_{3/2}$ electron occupation. For metallic Ir and Ir-alloys BR* is much smaller (≈ 3.5), indicating substantial quenching of $<LS>$, likely caused by the increased bandwidth induced by $5d$ hybridization [16,18]. However, orbital moment quenching $<LS>$ = 0 does not necessarily imply the absence of SO interaction in the Hamiltonian $H_{SO} = \zeta \times L \cdot S$, where $\zeta$ is the SOC constant/energy. When SOC dominates ($\zeta \gg 10Dq$), the Ir $5d$ states split into $5d_{5/2}$ and $5d_{3/2}$ states, with $L_2$ and $L_3$ absorption corresponding to transitions $2p_{1/2} \rightarrow 5d_{3/2}$ and $2p_{3/2} \rightarrow 5d_{3/2,5/2}$, respectively: Notably, $I_{L_2} = 0$ for $Ir^{4+}(5d^5)$, which is inconsistent with observations in iridates. Furthermore, the ground state is often well described by the pseudo-spin scenario. In strong SOC limit ($\zeta \leq 10Dq$), the octahedral CF splits the $5d$ states into $e_g$ and $t_{2g}$ states, with the $t_{2g}$ further split by SOC into a fully occupied $J_{eff}=3/2$ quadruplet and a half-filled $J_{eff}=1/2$ doublet. Note that in this case orbital mixing between the $e_g$ and $t_{2g}$ states is generally neglected.

Hydrostatic pressure $p$ offers a powerful mean to tune electronic bandwidth, octahedral distortions/rotations, and CF making it ideal for studying $<LS>$ and the SO coupled ground state in SIO. The layered iridates exhibit a robust $J_{eff}=1/2$ ground state with only moderate quenching of $<LS>$ under increasing $p$ [19,20]. However, with greater connectivity of $IrO_6$ octahedra occurring $e.g.$ when going from corner-sharing to edge- or face-sharing configurations, the reduced Ir-Ir distance facilitates molecular orbital formation and the breakdown of the $J_{eff}=1/2$ ground state [8]. Covalency-driven orbital mixing thus plays a critical role in shaping the pseudospin scenario in iridates, motivating detailed studies of SIO under $p$.

The availability of bulk SIO is limited by the impossibility of producing single crystals under ambient $p$ [21,22]. Nevertheless, single-crystalline SIO can be synthesized as epitaxial thin films, where the orthorhombic structure is stabilized by strain [23–27]. Using pulsed laser deposition, we have successfully grown SIO films on a water-soluble $Sr_4Al_2O_7$ sacrificial layer, enabling the preparation of structurally relaxed freestanding single-crystalline SIO films. This approach has allowed us to study the SO coupled ground state of bulk-like SIO. We performed Ir $L_{2,3}$ XAS under hydrostatic $p$, analyzing the branching ratio and $<LS>$ as function of $p$ at room temperature. The Ir $L_{2,3}$ XANES indicate a rather large value of $<LS>$ at ambient $p$ suggesting significant mixing of $J_{eff}=3/2$ and $e_g$ orbitals. With increasing $p$, $<LS>$ decreases consistent with the CF enhancement due to a $p$-induced shortening of the Ir-O bond-length. Despite the semi-metallic character of SIO, the $p$ induced changes in $<LS>$ are remarkably similar to those observed in the layered insulating iridates $Sr_2IrO_4$ and $Sr_3Ir_2O_7$, indicating a robust $J_{eff}=1/2$ ground state for SIO.

II. Experimental

Single crystalline SIO was produced by thin film deposition technique [28]. To this end, epitaxial SIO films with a thickness of about 120 nm were deposited by pulsed laser deposition on a water-soluble $Sr_4Al_2O_7$ thin film which was first grown epitaxially on a (001) $SrTiO_3$ substrate, see Supplemental Material [29]. The films were removed from the substrate and transferred to a thermal release tape, attached to the SIO surface before, by dissolving the water-soluble sacrificial layer. Finally, the SIO single-crystal like freestanding membranes were transferred to the diamond culets of the used diamond anvil cells by stamping the membranes onto the culet surface under the load of heat and weight [29]. Using membrane-driven partially perforated diamond anvil cells with 600 μm and 350 μm culet, hydrostatic $p$ could be continuously varied up to 50 GPa. The cells were filled with He gas and sealed with steel gaskets. $p$ was measured in-situ using ruby fluorescence. The XANES at the Ir $L_{2,3}$-edge was done at the beam line ID12 at the European Synchrotron Radiation Facility in Grenoble, France. Measurements were carried out using circular polarized light with normal beam incidence at room temperature using the HU-38 undulator and an energy-resolved detector. All the spectra were recorded using the partial fluorescence yield with two silicon drift detectors with Falcon X signal processor. Due to the low thickness of the film (120 nm), no reabsorption corrections were needed. Further information on data analysis are described in Ref. [29].

III. Results

A. Structural properties of $SrIrO_3$

The structural characterization of the SIO freestanding films was carried out by X-ray diffraction (XRD) and scanning transmission electron microscopy (STEM). Symmetric θ/2θ diffraction, shown in Fig. 1(a), confirm the (110) growth orientation of the freestanding film. The orthorhombic structure was further characterized by probing asymmetric reflections of the pseudo-cubic types <113> and <123>. Half-integer reflections, indicative of a doubling of the pseudo-cubic unit cell, were observed for every reflection except for the (312), as shown in Fig. 1(b). These observations allowed the deduction of the octahedral rotation pattern characteristic of the orthorhombic structure. For the out-of-plane film axis, only antiphase rotation (-) was observed, whereas for the in-plane axes both, in-phase (+) and antiphase (-) rotations were present. This pattern is consistent with in-plane twinning of the orthorhombic $c$-axis, expected due to epitaxial growth on the quadratic surface cells of $Sr_4Al_2O_7$ and $SrTiO_3$. According to Glazer notation, the octahedral tilt pattern is ($a^-a^-c^+$), identical to that of bulk orthorhombic (*Pbnm*) SIO [22,30].

Plan-view high-resolution (HR) STEM images and corresponding fast Fourier transforms (Fig. 1(c)) also confirm the orthorhombic structure with the $c$-axis aligned perpendicular to the surface normal. The integrated differential phase contrast (iDPC) STEM image (Fig. 1(d)) further reveal the antiphase rotation along the out-of-plane direction, evidenced by elongated oxygen atom columns (blue ellipses in Fig. 1(d)) resulting from antiphase-rotated octahedra. The antiphase rotation was estimated to (-) ≈ 8°. Lattice parameters deduced form X-ray diffraction (see also Ref. [29]) are $a$ = 5.61(1) Å, $b$ = 5.60(5) Å and $c$ = 7.91(6) Å. These parameters and rotation angles align well with those observed for the high-$p$ phase of SIO synthetized at 6 GPa ($a$ = 5.60 Å, $b$ = 5.57 Å, $c$ = 7.90 Å, (+) = 8.7°, and (-) = 12°) [31]. The slightly smaller lattice parameters and larger rotation angles are likely explained by the high-$p$ synthesis conditions [22].

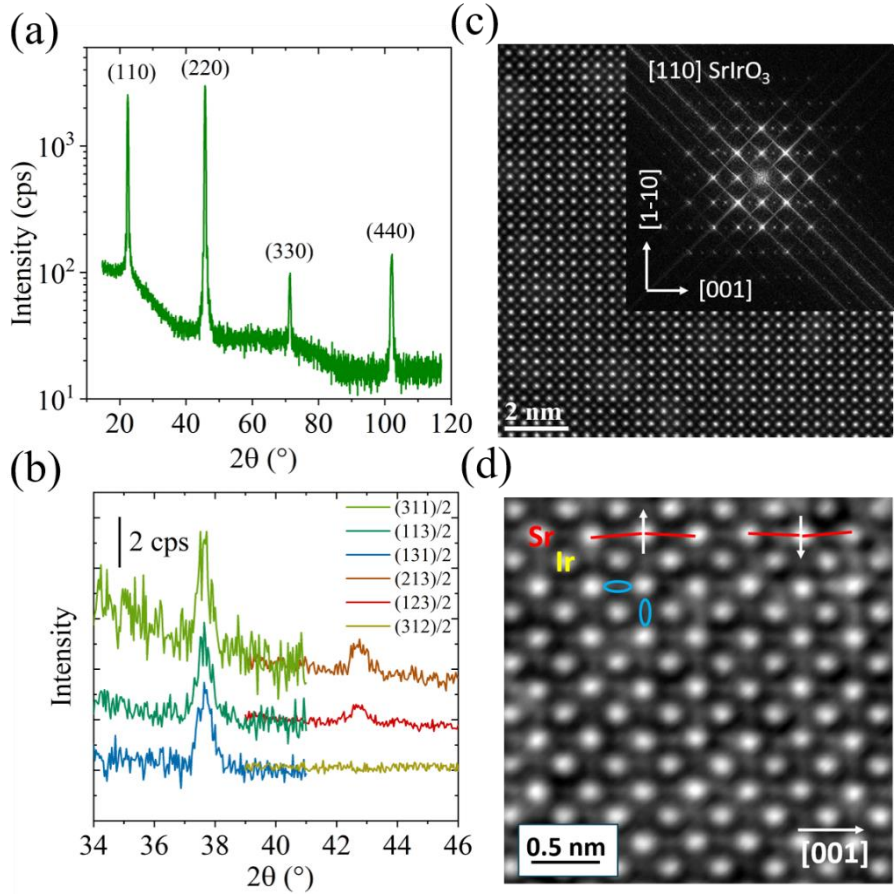

**FIG. 1.** Structural properties of SIO freestanding films. (a) Symmetric θ/2θ scan with respect to surface-normal, indicating pure (110) growth orientation. (b) The half-integer reflections of the pseudo-cubic <113> and <123> peaks. (312)/2 peak indicating (+) rotation for the out-of-plane direction is absent. The peaks are shifted vertically for clarity. The pseudo-cubic axis *a*, *b*, and *c* correspond to the orthorhombic [1-10], [001], and [110] direction, respectively. (c) Plan-view HR-STEM image and the corresponding fast Fourier transformation in inset. (d) iDPC-STEM image, showing untwinned domain with in-plane alignment of the c-axis and (-) rotation along the out-of-plane direction. Red lines and blue ellipses indicate Sr and O displacement, respectively.

B. X-ray absorption spectroscopy at the Ir $L_{2,3}$ edge

The *p*-dependence of <LS> in SIO was investigated by XAS. XANES spectra at the Ir $L_{2,3}$-edge were recorded under hydrostatic *p* ranging from 0.37 to 47.4 GPa at room temperature. The most notable feature in the Ir $L_2$ and $L_3$ XANES spectrum is the sharp WL corresponding to the $2p_{1/2} \rightarrow 5d_{3/2}$ and $2p_{3/2} \rightarrow 5d_{3/2,5/2}$ electronic transitions, respectively. Figure 2 shows the normalized intensity (see [29]) of the Ir $L_{2,3}$ XANES across the *p* range. The WL intensity of the Ir $L_3$ XANES decreases markedly with *p*, while that of the $L_2$ XANES increases only slightly. The extended X-ray absorption fine structure (EXAFS) oscillations observed after the WL shift to higher energies with increasing *p*, indicative of a shortening of the interatomic distances between Ir and the next nearest oxygen atoms, as expected under compression [32,33]. The energy shift of the EXAFS oscillations is similar at both edges ($L_3$ and $L_2$) and amounts to approximately 7.5 eV at *p* = 47.4 GPa. However, due to *p*-dependent variations in the background absorption, accurate determination of the EXAFS maximum was challenging. The energy $E_0$, corresponding to the maximum of the $L_{2,3}$ WL also shifts, albeit by a smaller amount (≈ 0.8 eV) compared to the EXAFS oscillations. Nevertheless, both features - WL and EXAFS - shift nearly linearly with *p*, indicating continuous changes in the electronic and local structural properties of the IrO$_6$ octahedra (see Fig. 3(a)).

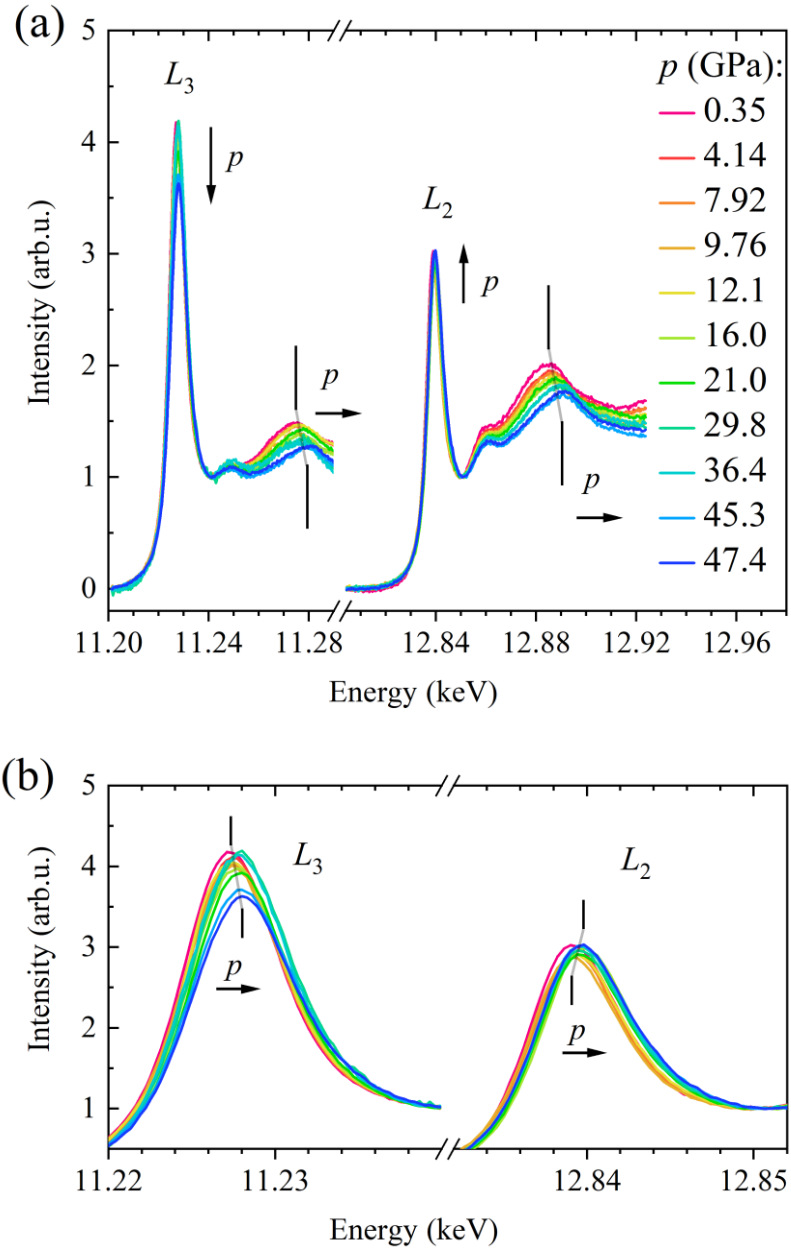

**FIG. 2.** (a) Normalized intensity of the Ir $L_3$ and $L_2$ XANES of SIO for different $p$ ranging from 0.35 – 47.4 GPa. Measurements were taken in normal beam incidence (90°) at $T$ = 300 K. With increasing $p$, the peak intensity of the $L_3$ decreases and of the $L_2$ increases, as indicated by the arrows. The EXAFS feature after the WL (indicated by bars) also shifts to higher energy with increasing $p$. (b) Enlarged view on the $L_3$ and $L_2$ WL to visualize $p$-dependence of the intensity and peak position. Color scale is same as above.

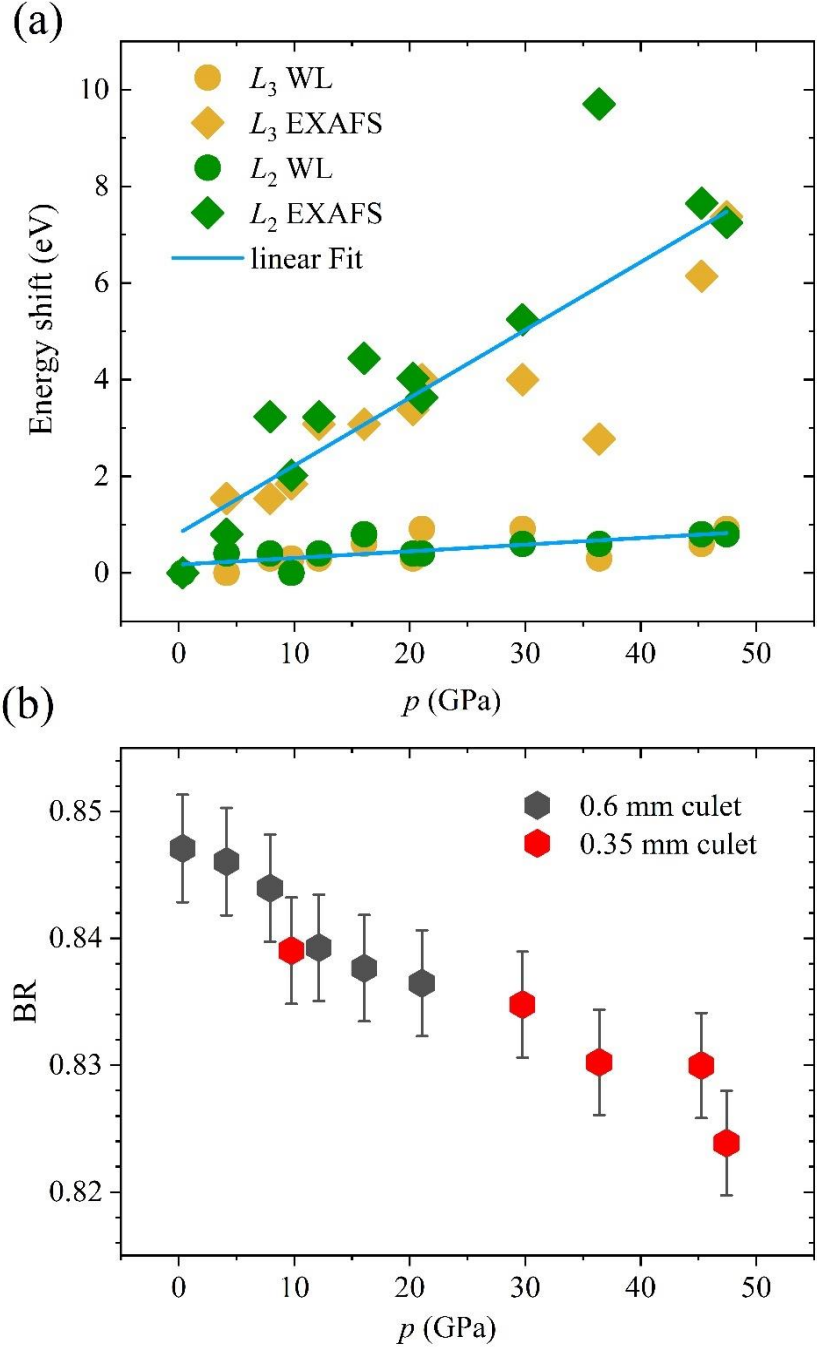

**FIG. 3.** (a) Energy shift of the peak position $E_0$ of the Ir $L_3$ and $L_2$ WL and EXAFS wiggles versus hydrostatic pressure $p$. The energy shifts nearly linear with $p$ and amount to about 0.8 eV and 7.5 eV for the WL and EXAFS features, respectively, at $p = 47$ GPa. (b) The branching ratio BR = $I_{L3}^{XAS}/(I_{L3}^{XAS} + I_{L2}^{XAS})$ versus $p$. Measurements were obtained from different diamond anvil cells, see legend. Error bars are related to WL area analysis.

Next, we quantify the effect of SOC on the 5$d$ states through the branching ratio of the WL integrals at the SO split absorption edges for the $p \rightarrow d$ transitions. The branching ratio is defined as BR = $I_{L3}^{XAS}/(I_{L3}^{XAS}+I_{L2}^{XAS})$, where $I_{L2,3}^{XAS}$ represent the integrated WL intensities, determined after subtraction of the background using an arctangent function [29]. This ratio is related to the expectation value of the angular part of the ground-state SO operator <$LS$> trough the SO sum rules: BR = 2/3 + <$LS$>/(3×$n_h$) [17]. As already suggested by Fig.2, $I_{L3}^{XAS}$ decreases while $I_{L2}^{XAS}$ increases with $p$. This trend is consistent with observations in $Sr_2IrO_4$ and $Sr_3Ir_2O_7$ [19,20]. The sum $I_{L3}^{XAS}+I_{L2}^{XAS}$, which is proportional to the number of Ir 5$d$ holes $n_h = n_h^{j=3/2}+ n_h^{j=5/2} \approx 5$, remains nearly constant with $p$ [29]. Therefore, the shift in spectral weight from $I_{L3}^{XAS}$ to $I_{L2}^{XAS}$ with increasing $p$ suggests a progressive mixing or hybridization of occupied $5d_{3/2}$ - and unoccupied $5d_{5/2}$ –states.

Fig. 3(b) shows the branching ratio as a function of $p$. At moderate $p < 5$ GPa, the BR ($\approx 0.846$) is comparable to that of compressively strained epitaxial SIO films (BR $\approx 0.844$), investigated in previous studies. The observed BR is significantly higher than the statistical value of 2/3 (<LS> = 0) indicating the presence of strong SOC effects. This BR is consistent with values reported for other Ir compounds such as $Sr_3Ir_2O_7$, $Sr_2IrO_4$, $IrO_2$, or $IrCl_3$ [16,19,20]. Notably, small deviations in BR can arise from differences in methods for the WL peak area extraction; however, the use of a consistent procedure ensures the robustness of our conclusions. With increasing $p$, the BR steadily decreases to 0.824 for $p = 47$ GPa, indicating a continuous reduction in <LS>. Importantly, this decrease is gradual, with no abrupt changes or drops indicative of orbital momentum quenching.

IV. Discussion

In perovskite iridates, the SOC energy of the 5$d$ electrons ($\zeta$) competes with the CF energy. The large spatial extension of the 5$d$-orbitals, results in large 10$Dq$ ($\approx 3$ eV) [34,35], which usually results in a low spin S = ½ ground state in the limit of weak SOC. However, strong SOC of the Ir ($\zeta = 0.45$ eV) [38–40], splits the $t_{2g}$ into a fully occupied $J_{eff} = 3/2$ quadruplet and a half-filled $J_{eff} = 1/2$ doublet. In such a ground state, the contribution of holes to the expectation value <LS> = -1/2 [$j_{eff}(j_{eff}+1)$-$l_{eff}(l_{eff}+1)$-$s(s+1)$] is given by the sum of the spin-orbit values <LS>$_{Jeff=3/2}$ (= 0) and <LS>$_{Jeff=1/2}$ (= 1), resulting in a BR = 0.73, which is significantly smaller than what we observed (Fig. 3a). In this scenario, only the SOC-induced mixing within the $t_{2g}$ manifold is considered ($\zeta < 10Dq$ - see also [29]). However, when $\zeta$ becomes comparable to 10$Dq$, the mixing between $t_{2g}$ and $e_g$ states in the fully diagonalized SOC Hamiltonian scales with $\zeta/(\zeta/2+10Dq)$ allowing up to 20% occupancy of orbitals that are typically unoccupied [41]. Configuration interaction (CI) calculations for an $Ir^{4+}$ ion in an octahedral CF also reveal significant mixing between the $J_{eff} = 3/2$ and $e_g$ states [42]. In addition to the contributions of the $J_{eff} = 1/2$ state to <LS>, orbital mixing between the symmetry equivalent $J_{eff}=3/2$ and $e_g$ states also results in a contribution of $e_g$ states to <LS>. The subsequent increase of <LS> follows the scaling relation $\zeta/10Dq$. For example, with $\zeta = 0.3$ eV and 10$Dq = 3$ eV the calculated total <LS> aligns closely with the value inferred from the experimental BR [42,43].

The large BR and $p$-induced changes in the Ir $L_{2,3}$ XANES of SIO reported here therefore strongly suggest the presence of orbital mixing of $J_{eff}=3/2$ and $e_g$ states. It is important to note that noncubic structural distortions, such as tetragonal CF splitting $\Delta_t$ can also influence the BR and the electronic ground state when it becomes comparable to $\zeta$ [6]. For comparison in the layered iridates $Sr_2IrO_4$ or $Sr_3Ir_2O_7$, $\Delta_t$ amounts to 1.6 eV and 1.9 eV [35], respectively, exceeding considerably $\zeta$.

With increasing $p$, the BR progressively decreases to 0.824 for $p = 47$ GPa, reflecting a continuous reduction of <LS>. This is illustrated in Fig. 4(a), where we have extracted the $p$ dependence of <LS> from the BR. The expectation value <LS> decreases by approximately 15% and signals a significant reduction in orbital angular momentum. Under increasing $p$, the Ir-O bond length is expected to shorten. For example, in $Sr_2IrO_4$ the $a$-lattice parameter contracts by about -0.15 %/GPa, leading to an estimated 50% enhancement of the CF strength at $p = 70$ GPa [19].

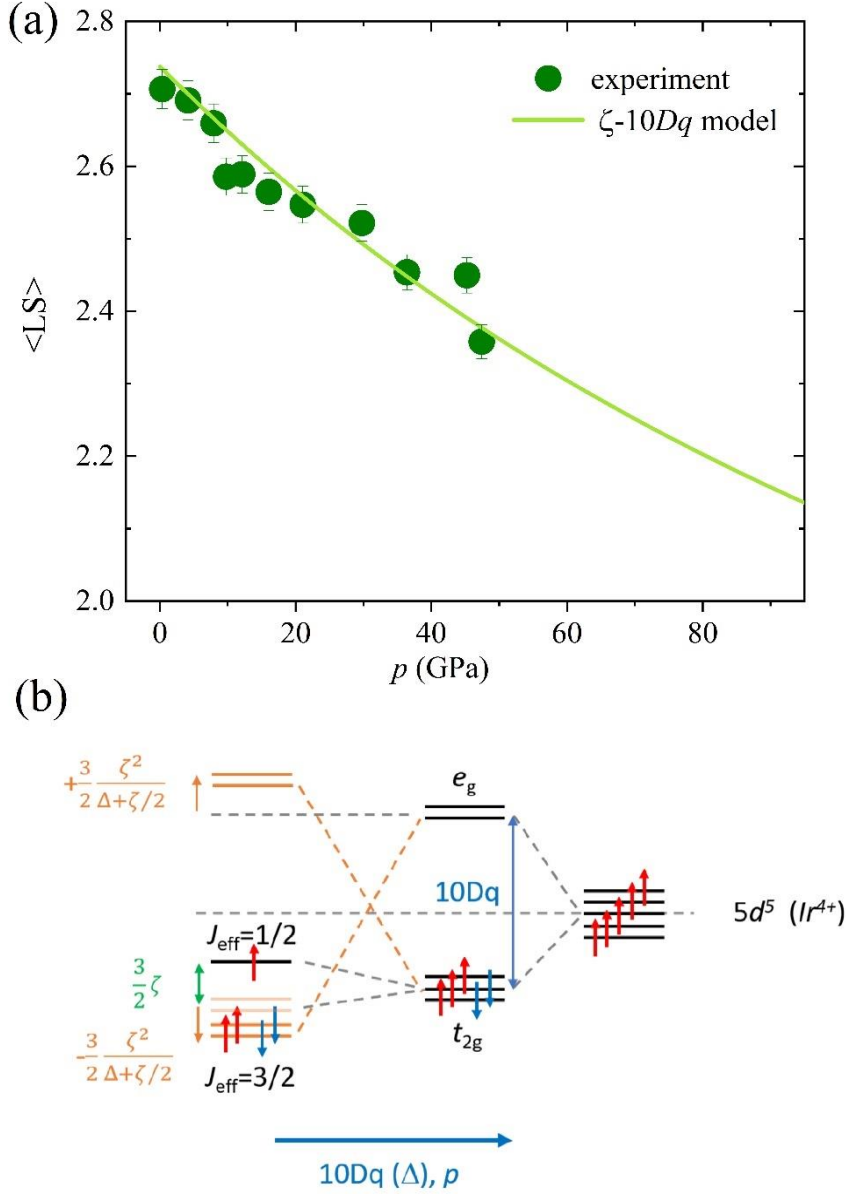

**FIG. 4.** (a) Effective SOC as deduced from the BR of SrIrO$_3$. Error bars are related to WL area analysis. The solid line indicates calculated values of <LS> based on the $\zeta$-10$Dq$ model of Donnerer et al. [20], see [29]. (b) Energy level scheme of the Ir5$d$ states split by CF and SOC. The 5$d$ states are split by the octahedral CF 10$Dq$ (right) and are than further split by SOC $\zeta$ (left) into $J_{eff}$=1/2 and $J_{eff}$=3/2 states. In the limit of $\zeta \leq 10Dq$, $e_g$-$t_{2g}$ orbital mixing results in an energy shift of the $J_{eff}$=3/2 ($e_g$) states by - (+) 3/2×$\zeta^2$/(10$Dq$+$\zeta$/2) [41]. With increasing pressure $p$, 10$Dq$ increases yielding a decrease of <LS>.

To probe the $p$ dependence of <LS> in relation to the $p$-induced increase of 10$Dq$, we employed the $\zeta$-10$Dq$ model, proposed by Donnerer et al. [20]. This single-ion model incorporates SOC and CF, with a complete diagonalization of the Hamiltonian that fully accounts for the mixing between the $t_{2g}$ and $e_g$ states. The model requires only one adjustable parameter, $\zeta$/10$Dq$, to calculate <LS>, see [20,29]. Following the literature, we used $\zeta$ = 0.3 eV [42], while 10$Dq$ at $p \approx 0$ was adjusted to 1.7 eV to reproduce the experimental value of <LS> = 2.7. This CF strength is consistent with first-principle calculations for SIO [44], and is distinctly smaller than experimental values reported for Sr$_2$IrO$_4$ (3.8 eV) and Sr$_3$Ir$_2$O$_7$ (3.55 eV) [35]. The lower 10$Dq$ for SIO compared to layered iridates can be attributed to its metallic nature and the absence of Mott-splitting in the $J_{eff}$=1/2 state. In Sr$_2$IrO$_4$, the Coulomb repulsion splits the $J_{eff}$=1/2 doublet by about 2 eV lowering the occupied $J_{eff}$=1/2 state by about 1 eV and thereby increasing the effective $e_g$-$t_{2g}$ splitting, *i. e*, $10Dq$ by a similar amount [45,46]. This explanation is supported by the comparable Ir-O bond distances in the rigid $IrO_6$ octahedra across these materials [31,38], while the electronic bandwidth near the Fermi energy, which is similar for SIO and $Sr_2IrO_4$ [27], is unlikely to significantly affect $10Dq$. The *p*-induced decrease in <LS> is therefore most likely driven by an increase in the CF due to the hydrostatic compression of SIO and the disentanglement of $J_{eff}$ =3/2 and $e_g$ states with increasing *p*.

Further experimental support for this scenario is also provided by the observed upshift of $E_0$. For *p* = 47 GPa, the $E_0$ positions of the $L_3$ and $L_2$ WLs shift by about 0.8 eV to higher energies, consistent with observations for $Sr_3Ir_2O_7$ under similar conditions [20]. Since the number of 5*d* holes remains nearly constant, varying by less than 5%, the shift in $E_0$ is most likely driven by CF effects and orbital mixing. According to XANES selection rules, the mixing of $J_{eff}$=3/2 and $e_g$ states allows transitions to empty $e_g$ states at both $L_2$ and $L_3$ edges, which are forbidden for the $L_2$ edge in the absence of orbital mixing. An increase of CF by $\delta 10Dq$ raises the energy of the $e_g$ states, leading to an upshift of the $L_2$ and $L_3$ WLs of similar magnitude. At *p* = 47 GPa, CF is expected to increase from 1.7 eV to 2.3 eV [29]. Albeit smaller, the calculated shift is on the same order of magnitude as the observed one, further supporting the orbital-mixing scenario. Additionally, tight-binding calculations predict an upshift of the $J_{eff}$=3/2 state with increasing $10Dq$ [41]. A corresponding energy level diagram illustrating this is provided in Fig. 4(b).

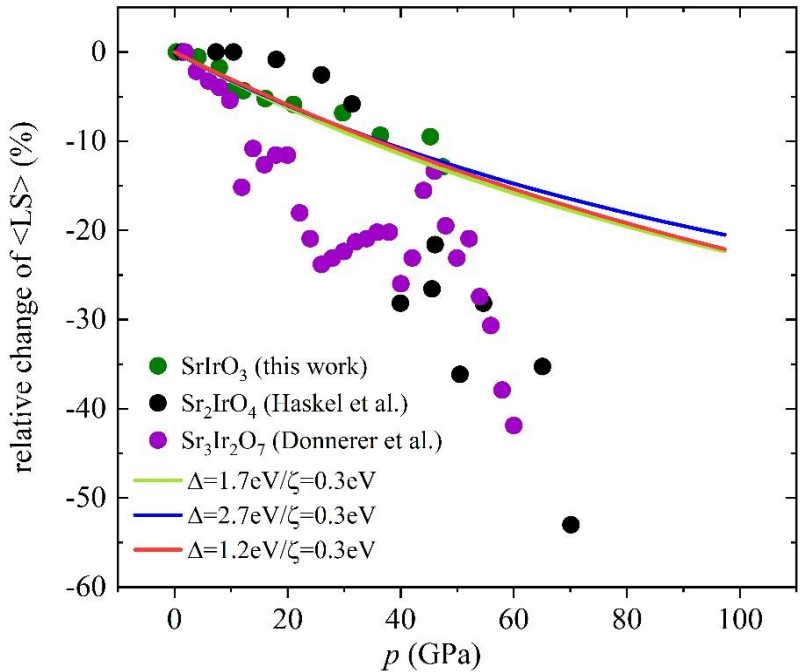

**FIG. 5.** Relative change of <LS> for the perovskite-related iridates $SrIrO_3$, $Sr_2IrO_4$, and $Sr_3Ir_2O_7$. Additional data were taken from Ref. [19,20]. Solid lines indicate simulations from $\zeta$-$10Dq$ model for $\zeta$ = 0.3 eV and $10Dq$ ($\Delta$) = 1.2 eV, 1.7 eV, and 2.7 eV.

To gain further insights on the influence of CF strength on $\langle LS \rangle$, it is instructive to compare the *p*-induced changes of the Ir $L_2$ and $L_3$ XANES of SIO, $Sr_2IrO_4$ and $Sr_3Ir_2O_7$ that exhibit notable similarities up to about 50 GPa. Specifically, $I_{L_3}^{XAS}$ decreases while $I_{L_2}^{XAS}$ increases with *p*. Figure 5 shows relative change of $\langle LS \rangle$ as derived from the BR as function of *p* for SIO compared to reported data [19,20] for the layered iridates $Sr_2IrO_4$ and $Sr_3Ir_2O_7$. The figure also includes simulations of $\langle LS \rangle$ for $\zeta$ = 0.3 eV and $10Dq$ = 1.2, 1.7 and 2.7 eV, corresponding to ±20% variations from the extracted ambient $p$ value for SIO. In the moderate $p$-regime ($p < 50$ GPa), these simulations clearly suggest only minor dependence of the relative $p$-induced decrease of ⟨$LS$⟩ on the strength of cubic CF 10$Dq$.

Although experimental BR and ⟨$LS$⟩ are reasonably well described by $p$-induced changes of the cubic CF 10$Dq$ and resulting orbital mixing of $e_g$ and $t_{2g}$ states in the low-$p$ regime [19,20], significant deviations emerge for the layered iridates at higher $p$. For $Sr_2IrO_4$ deviations are evident above 50 GPa, and for $Sr_3Ir_2O_7$ deviations are already apparent for 10 GPa. The simulations indicate that the larger 10$Dq$ values of the layered iridates, $e. g.$, 3.8 eV and 3.55 eV for insulating $Sr_2IrO_4$ and $Sr_3Ir_2O_7$, respectively, or 2.9 eV for non-metallic $BaIrO_3$ [42], are insufficient to account for these deviations. Instead, the observed discrepancies are more likely attributed to tetragonal distortion and noncubic CF effects, which can contribute to the quenching of ⟨$LS$⟩. Additionally, a structural phase transition in $Sr_3Ir_2O_7$ around 53GPa [47] could influence both the CF splitting and the BR. For $p > 50$ GPa, enhanced hybridization effects and bandwidth-driven orbital mixing may also become significant [19,47]. Unfortunately, the diamond anvil cells used in our experiment did not allow measurements for $p > 50$ GPa, leaving the high-$p$ behavior of SIO unexplored.

We conclude that the robustness of ⟨$LS$⟩ in SIO in the investigated $p$-range can likely be attributed to the absence of significant noncubic CF and of hybridization effects. Compounds with face-sharing (edge-sharing) $IrO_6$ octahedra and short $d_{Ir-Ir}$ ≈ 2.3 Å (≈ 2.8 Å), compared to that of corner-sharing case of SIO (≈ 3.96 Å), exhibit a covalency-driven collapse of the $J_{eff}$=1/2 state [8]. For metallic Ir ($d_{Ir-Ir}$ ≈ 2.6 Å), the BR at ambient conditions is close to the statistical value [16], indicating strong quenching of ⟨$LS$⟩. In contrast, the narrow electronic bandwidth and relatively large Ir-Ir distance ($d_{Ir-Ir}$ ≈ 3.96 Å) in SIO minimize direct overlap of neighboring 5$d$ orbitals, effectively preventing the formation of molecular orbitals by covalency and preserving the $J_{eff}$=1/2 pseudospin state.

V. Summary

In summary, the synthesis of orthorhombic perovskite SIO as freestanding thin-film material enabled a systematic investigation of the effective SOC in the corner sharing $IrO_6$ octahedra under hydrostatic $p$ and an evaluation of the $J_{eff}$=1/2 ground-state scenario. The angular part of the SO operator expectation value <$LS$> was derived from the BR of Ir $L_{2,3}$ XANES. At ambient $p$, the BR is relatively large (0.85) indicating a significant orbital mixing between the occupied $J_{eff}$=3/2 pseudospin ($t_{2g}$-related) states and unoccupied $e_g$ states. Upon increasing $p$, <$LS$> decreases by about 15% at $p ≈ 50$ GPa. This $p$-induced reduction in the BR and <$LS$> reflects a suppression of the $e_g$-$t_{2g}$ orbital mixing, consistent with an increase in the cubic CF parameter 10$Dq$ caused by the shortening of the Ir-O bond-length under compression. The cubic CF parameter for SIO, derived from the $\zeta$-10$Dq$ model, is 1.7 eV at ambient $p$, which is notably smaller than the values reported for the layered iridates. This difference is likely attributed to SIO´s metallic nature and the absence of a Mott-driven splitting of the $J_{eff}$=1/2 doublet at ambient $p$. Despite this semi-metallic character, the SOC-driven $J_{eff}$=1/2 ground state is found to be remarkably robust across the investigated $p$-range. The results provide compelling evidence for the role of orbital mixing and the interplay of SOC and CF effects in shaping the electronic structure of SIO.

**Acknowledgements**

We thank R. Eder for fruitful discussion and is grateful to R. Thelen and the Karlsruhe Nano-Micro Facility (KNMFi) for technical support. A. K. J. acknowledges financial support from the European Union's Framework Programme for Research and Innovation, Horizon 2020, under the Marie Skłodowska-Curie grant agreement No. 847471 (QUSTEC).